\documentclass[notitlepage,prd,superscriptaddress,onecolumn,nofootinbib,showpacs]{revtex4-1}
\usepackage{amsfonts,amssymb,amsmath}
\usepackage{mathrsfs}
\usepackage{color}
\usepackage{graphicx}
\usepackage{dcolumn}
\usepackage{bm}

\usepackage{pslatex}
\usepackage[T1]{fontenc}
\usepackage[latin1]{inputenc}
\usepackage{graphicx}
\usepackage{epsfig}
\usepackage{longtable}
\usepackage{float}
\usepackage{calc}
\usepackage{ifthen}

{
{
{
\newcommand{\bea}{\begin{eqnarray}}
\newcommand{\eea}{\end{eqnarray}}

\newcommand{\nc}{\newcommand}
\nc{\renc}{\renewcommand}
\nc{\eqs}[2]{\mbox{Eqs.~(\ref{#1},\,\ref{#2})}}
\nc{\eq}[1]{\mbox{Eq.~(\ref{#1})}}
\nc{\figs}[2]{\mbox{Figs.~(\ref{#1},\,\ref{#2})}}
\nc{\fig}[1]{\mbox{Fig~.(\ref{#1})}}
\nc{\be}[1]{\begin{equation} \mbox{$\label{#1}$}}
\nc{\ee}{\vspace{0.1cm}\end{equation}}

\newcommand{\bean}{\begin{eqnarray*}}
\newcommand{\eean}{\end{eqnarray*}}

\begin{document}

\title{Freeze-In Dark Matter from a sub-Higgs Mass Clockwork Sector via the Higgs Portal}
\author{Jinsu Kim}
\email{kimjinsu@kias.re.kr}
\affiliation{Quantum Universe Center, Korea Institute for Advanced Study, Seoul 02455, Korea}
\author{John McDonald}
\email{j.mcdonald@lancaster.ac.uk}
\affiliation{Department of Physics, Lancaster University, Lancaster LA1 4YB, UK}

\begin{abstract}

The clockwork mechanism allows extremely weak interactions and small mass scales to be understood in terms of the structure of a theory. A natural application of the clockwork mechanism is to the freeze-in mechanism for dark matter production. Here we consider a Higgs portal freeze-in dark matter model based on a scalar clockwork sector with a mass scale which is less than the Higgs boson mass. The dark matter scalar is the lightest scalar of the clockwork sector. Freeze-in dark matter is produced by the decay of thermal Higgs bosons to the clockwork dark matter scalars. We show that the mass of the dark matter scalar is typically in the 1-10 keV range and may be warm enough to have an observable effect on perturbation growth and Lyman-$\alpha$ observations. Clockwork Higgs portal freeze-in models have a potentially observable collider phenomenology, with the Higgs boson decaying to missing energy in the form of pairs of long-lived clockwork sector scalars, plus a distribution of different numbers of quark and lepton particle-antiparticle pairs. The branching ratio to different numbers of quark and lepton pairs is determined by the clockwork sector parameters (the number of clockwork scalars $N$ and the clockwork charge $q$), which could therefore be determined experimentally if such Higgs decay modes are observed. In the case of a minimal Standard Model observable sector, the combination of nucleosynthesis and Lyman-$\alpha$ constraints is likely to exclude on-shell Higgs decays to clockwork scalars, although off-shell Higgs decays would still be possible. On-shell Higgs decays to clockwork scalars can be consistent with cosmological constraints in simple extensions of the Standard Model with light singlet scalars.

\end{abstract}

\maketitle

\section{Introduction}
\label{sec:intro}

The clockwork mechanism \cite{choi,cl1,giudice} is a way to understand very small masses and couplings in terms of the structure of a theory, rather than in terms of symmetries\footnote{
See \cite{CWgeneral} and \cite{CWgauged} for frameworks that generalize the approach of \cite{choi,cl1,giudice}.
}. It can be motivated by the principle that in any theory with a characteristic mass scale, the natural values for the mass terms and couplings are either zero or naturally large. Couplings which are extremely small compared to one or masses which are extremely small compared to the characteristic mass scale of the theory should then be explained by a particular assignment of masses and couplings, which are either zero or of a natural magnitude.

A range of applications of the clockwork mechanism has been proposed, including neutrino masses through the seesaw mechanism \cite{CWDM, CWneutrino}, muon $g-2$ \cite{CWmuong2}, axions \cite{CWaxions}, dark matter \cite{CWDM}, composite Higgs \cite{CWComposite}, the Weak Gravity Conjecture \cite{wgc} and inflation \cite{CWinflation}. (See \cite{craig} and \cite{reply} for further discussion of the clockwork mechanism.) A particularly natural application of the clockwork mechanism is to freeze-in dark matter \cite{jmfr,hall}. (For a review of the freeze-in mechanism, see \cite{tenkanen}.) Freeze-in dark matter requires a dark matter particle which has a very small mass compared to the weak interaction mass scale $O(m_{W})$ and, in order to keep the dark matter particles out of thermal equilibrium\footnote{
In the case of sizeable dark matter self-interactions, the dark matter particles may thermalize themselves, and the dark matter abundance is governed by the so-called dark freeze-out mechanism \cite{Heikinheimo:2016yds,Heikinheimo:2017ofk,Enqvist:2017kzh}.
}, very weak interactions with the Standard Model (SM) sector particles. In particular, for the case of freeze-in dark matter via the Higgs portal, there is no symmetry that can suppress the Higgs portal interaction. This suggests that a structural explanation for the suppression of the Higgs portal coupling may be necessary.

In the approach of \cite{cl1}, the clockwork sector is viewed as the low-energy effective theory of a theory consisting of $N+1$ scalars $\phi_{j}$ with a global $U(1)^{N+1}$ symmetry, which is spontaneously broken at a scale $f$. The clockwork sector of the effective theory corresponds to the Goldstone bosons, $\pi_{j}$, $j = 0, ... ,N$, of the spontaneously broken global symmetry. The $U(1)^{N+1}$ global symmetry is further broken by explicit symmetry breaking mass and coupling terms (assumed to originate from a UV completion). These terms leave a single residual spontaneously broken $U(1)$ symmetry. In this case there is a single massless Goldstone boson, $a_{0}$. The other $N$ clockwork scalars, $a_{1}, ... ,a_{N}$, have a spectrum masses, which are all of the order of the characteristic mass scale of the clockwork sector. The $a_{0}$ scalar is completely decoupled from the potential.

The residual spontaneously broken $U(1)$ is explicitly broken by the coupling of the clockwork sector to the SM sector, which also gives a mass to the $a_{0}$ scalar.
A key assumption of clockwork models is that the coupling of the clockwork sector to the SM is only via the $\pi_{N}$ scalar. The $a_{0}$ scalar contributes a very small component to $\pi_{N}$, therefore its mass will be much smaller than the characteristic scale of the clockwork sector and its couplings to the SM will also be highly suppressed.

The restriction of the coupling of the clockwork sector to the SM sector to be via the $\pi_{N}$ scalar is assumed to be a feature of the underlying dynamics of the model. It is analogous to the assumption in 5-dimensional brane models that the SM fields are restricted to a single point in the extra-dimension. The clockwork sector mass and coupling terms can be understood in terms of a discrete extra-dimension, with the index $j$ of the $\pi_{j}$ playing the role of the 5th-dimensional coordinate \cite{cl1, giudice}. Restriction of the SM field to a brane at $j = N$ then restricts the coupling of the SM fields to be only to the $\pi_{N}$ scalar.

In the case of freeze-in models, in order to ensure that the freeze-in scalars have a number density that is much less than their thermal equilibrium density, the model typically has to have suppressed non-renormalizable derivative interactions of the $\pi_{j}$ scalars which are otherwise allowed by the $U(1)^{N+1}$ symmetry of the UV theory. In order to have a clockwork model that is compatible with this requirement, we will consider here a minimal clockwork sector corresponding to a renormalizable sector of scalars $\pi_{j}$ with a shift symmetry. In this case we can consider the SM sector and the clockwork sector to have a common origin in a single UV complete theory with a large characteristic energy cut-off $\Lambda$, below which the effective theory is renormalizable up to terms which are suppressed by powers of $\Lambda$.

A renormalizable clockwork sector can also be motivated by naturalness of the electroweak scale. It has been proposed that the SM is natural, in the sense that there are no large contributions to the Higgs mass due to quantum corrections, if there is no field with a mass between the weak scale and the Planck scale \cite{noscale}\footnote{See also the discussion in \cite{agrav}.}. In this case $\Lambda \gtrsim M_{Pl}$ is expected, with the theory becoming  renormalizable below the Planck scale.  

In \cite{chp1} we proposed a freeze-in model for scalar dark matter based on a TeV-scale scalar clockwork sector which couples to the SM via a Higgs portal interaction. In this case the dark matter scalars are produced by decay of heavy clockwork scalars which are in thermal equilibrium. Here we will consider the freeze-in cosmology of a sub-Higgs mass scalar clockwork sector which couples to the SM sector via the Higgs portal. In this case freeze-in will occur primarily via the decay of thermal bath Higgs bosons via $h \rightarrow \hat{a}_{k} + \hat{a}_{0}$, where the lightest clockwork scalar $\hat{a}_{0}$ is the dark matter scalar. We will show that, in general,  the dark matter scalars can be warm enough to have a significant effect on perturbation growth and Lyman-$\alpha$ observations. We will also show that clockwork Higgs portal freeze-in models have a potentially observable Higgs decay phenomenology, with $h$ promptly decaying to pairs of next-to-lightest clockwork scalars, $\hat{a}_{1}$, which will escape the detector, plus a distribution of quark and lepton particle-antiparticle pairs. 

Our main results for Higgs decay phenomenology can be summarized as follows. We find that in the case where the clockwork sector is coupled to the minimal SM, on-shell Higgs decays to clockwork scalars are excluded by nucleosynthesis and Lyman-$\alpha$ constraints.  On-shell Higgs decays to clockwork scalars are, however,  possible in simple extensions of the SM with light singlet scalars \cite{ram}, which can allow the clockwork scalars to decay before nucleosynthesis. Off-shell Higgs decays to clockwork scalars are possible in all models.

The paper is organized as follows. In Section \ref{sec:CWsector} we introduce the minimal renormalizable clockwork sector and its Higgs portal coupling to the SM sector.
In Section \ref{sec:CWfreezeinDM} we determine the conditions on the clockwork sector for $\hat{a}_{0}$ scalars from freeze-in to explain the observed dark matter density. 
In Section \ref{sec:Constraints} we consider the thermal history of the model and determine the conditions on the scale $\Lambda$ and the reheating temperature $T_{R}$ necessary to keep the freeze-in scalars out of thermal equilibrium.
In Section \ref{sec:Collider} we comment on the possible collider phenomenology of the model. In Section \ref{sec:NWDM} we discuss nucleosynthesis and warm dark matter constraints on the model.
In Section \ref{sec:Conclusion} we present our conclusions.

\vspace{0.2cm}

\section{A minimal renormalizable scalar clockwork sector}
\label{sec:CWsector}
In the following we will consider a $Z_2$ symmetric renormalizable scalar clockwork sector characterized by a mass scale of the order of the weak interaction mass scale, $O(m_{W})$. This represents a minimal construction of a scalar clockwork model. This construction is also compatible with naturalness of the weak scale, in the sense proposed in \cite{noscale}. We will also consider the most important non-renormalizable interactions which are consistent with the shift symmetry of the clockwork sector. These are due to derivative interactions suppressed by the fundamental cut-off scale $\Lambda$ of the renormalizable effective theory. 

The minimal $Z_{2}$-symmetric renormalizable potential of the clockwork sector is given by
\begin{align}\label{eqn:1}
	V_{CS} = \frac{1}{2} \sum_{j = 0}^{N-1} m_{j}^{2} (\pi_{j} - q \pi_{j+1})^{2} + \frac{1}{4}   \sum_{j = 0}^{N-1} \lambda_{\pi_{j}} (\pi_{j} - q \pi_{j+1})^{4} \,.
\end{align}
This is symmetric under $\pi_{j} \leftrightarrow -\pi_{j}$, where all $\pi_{j}$ are simultaneously transformed.
The first term corresponds to mass terms between the $\pi_{j}$ that are either zero or of the order of a common mass scale, in our case of the order of the weak scale. The couplings $\lambda_{\pi_{j}}$ are all considered to be naturally large (in the sense of not extremely small compared to 1). Following \cite{giudice}, we will simplify to the case where $m_{j} = m = O(m_{W})$ and $\lambda_{\pi_{j}} = \lambda_{\pi} = O(1)$ for all $j$. The potential \eqref{eqn:1} allows a single shift symmetry under which the fields simultaneously transform according to
\begin{align}\label{eqn:2}
	\pi_{j} \rightarrow \pi_{j} + \frac{\lambda}{q^{j}}\,,
\end{align}
where $\lambda$ is an arbitrary constant.
Therefore the potential has a flat direction and a corresponding massless scalar. On diagonalizing the mass matrix, the mass eigenstate scalars are $a_{i}$, $i = 0, ... ,N$. $a_{0}$ is at this stage a massless scalar, corresponding to the flat direction of the potential.

The mass eigenstate scalars are related to the $\pi_{j}$ by \cite{cl1,giudice}
\begin{align}\label{eqn:3}
	\pi_{j} = O_{ji}a_{i}\,,
\end{align}
where
\begin{align}\label{eqn:4}
	O_{j0} = \frac{\tilde{N}_{0}}{q^{j}}
	\,,\qquad
	O_{jk} = 
	\tilde{N}_{k} \left[ q \sin \left(\frac{jk\pi}{N+1}\right) - \sin \left( \frac{ \left(j + 1 \right) k \pi}{N+1} \right) \right]\,.
\end{align}
Here and in the following $i,\,j = 0, \cdots , N$ and $k = 1, \cdots , N$. $\tilde{N}_{0}$ and $\tilde{N}_{k}$ are given by
\begin{align}\label{eqn:5}
	\tilde{N}_{0} = 
	\sqrt{ \frac{q^2 - 1}{q^2 - q^{-2N}} }
	\,,\qquad
	\tilde{N}_{k} = \sqrt{\frac{2}{\left(N+1\right) \lambda_{k}}}
	\,,
\end{align}
where
\begin{align}\label{eqn:6}
	\lambda_{k} = q^2 + 1 - 2 q \cos \left(\frac{k \pi}{N+1} \right)\,.
\end{align}
The masses of the eigenstate scalars are
\begin{align}\label{eqn:7}
	m_{a_{0}}^2 = 0\,,\qquad
	m_{a_{k}}^2 = \lambda_{k} m^2
	\,.
\end{align}
In particular, for large $N$ we have $m_{a_{1}} \approx (q-1) m$ and $m_{a_{N}} \approx (q + 1) m$. 

In keeping with the construction of clockwork models, we assume that the shift symmetry is broken only by interactions of the $\pi_{N}$ scalar with itself or with the SM sector. The only renormalizable interaction of $\pi_{N}$ with the SM is via the Higgs bilinear, $H^{\dagger}H$. The renormalizable shift symmetry-breaking terms are therefore given by
\begin{align}\label{eqn:8}
	V_{\pi_{N}}  =  \frac{m_{N}^{2}}{2} \pi_{N}^{2} + \lambda_{h \pi_{N}}  \pi_{N}^{2} |H|^{2} + \frac{\lambda_{\pi_{N}}}{4} \pi_{N}^{4} \,,
\end{align}
where $m_{N}$ is assumed to be of the order of the clockwork scalar mass terms $m_{j}$ and smaller than the SM Higgs mass, and $\lambda_{h \pi_{N}}$ and $\lambda_{\pi_{N}}$ are assumed to be of a similar order of magnitude.  These terms will give the resulting lightest scalar of the clockwork sector, which we denote by $\hat{a}_{0}$, a very small mass compared to the weak scale and an extremely weak Higgs portal interaction with the SM sector.

In addition, if we consider the renormalizable theory to be an effective theory below a cut-off $\Lambda$, then there can be non-renormalizable interactions, which respect the $Z_2$ and shift symmetry of the clockwork sector, of the form
\begin{align}\label{eqn:9}
	{\cal L}_{NR} = \frac{1}{\Lambda^4} \sum_{i,j,k,l = 0}^{N} \gamma_{ijkl} \partial_{\mu} \pi_{i} \partial^{\mu}\pi_{j} \partial_{\nu} \pi_{k} \partial^{\nu} \pi_{l} \,,
\end{align}
where $\gamma_{ijkl}$ are $O(1)$ constants. For successful freeze-in, the scale $\Lambda$ and the reheating temperature after inflation $T_{R}$ must be such that the freeze-in dark matter scalars $\hat{a}_{0}$ do not acquire a thermal equilibrium number density due to these interactions.

\section{Clockwork scalar dark matter from freeze-in} 
\label{sec:CWfreezeinDM}

\subsection{Mass eigenstates and mixing due to the Higgs portal} 
The mass terms from Eq.~\eqref{eqn:1} and Eq.~\eqref{eqn:8} are
\begin{align}\label{eqn:10}
	\frac{\overline{m}_{N}^{2} \pi_{N}^{2}}{2} +  \frac{\sum_{k=1}^{N} m_{a_{k}}^{2} a_{k}^{2}}{2} =  
	\frac{\overline{m}_{N}^{2} (\sum_{j = 0}^{N} O_{Nj} a_{j})^{2}}{2}
	+ \frac{\sum_{k=1}^{N}m_{a_{k}}^{2} a_{k}^{2}}{2}\,,
\end{align}
where
\begin{align}\label{eqn:11}
	\overline{m}_{N}^{2} = m_{N}^{2} + \lambda_{h \pi_{N}} v^2\,,
\end{align}
with $v$ being the Higgs vacuum expectation value.
To have a case that can be diagonalized analytically, we will consider the limit where $O_{Nk}^{2}$ is small compared to 1 and $m_{a_{k}}^{2} \gtrsim \overline{m}_{N}^{2}$ for all $k$. It is easy to show, by maximizing all the factors in $O_{Nk}$, that in general $O_{Nk}^{2} < 2(q/(q-1))^{2}/(N+1)$. In the models discussed later we will consider $N \geq 10$, $q \geq 2$ and $ m_{a_{k}}^{2} \geq \overline{m}_{N}^{2}$, so the conditions for the approximate diagonalization to be valid will be satisfied. In this case the mass matrix becomes 
\begin{align}\label{eqn:12}
	\frac{\overline{m}_{N}^{2}}{2} (O_{N0}^{2} a_{0}^{2} + 2 O_{N0} \sum_{k = 1}^{N} O_{Nk} a_{0} a_{k})  +  \frac{\sum_{k = 1}^{N}  m_{a_{k}}^{2} a_{k}^{2} }{2}
	\,.
\end{align}
The mass eigenstates are then
\begin{align}\label{eqn:13}
	\hat{a}_{0} \approx a_{0} - \sum_{k = 1}^{N} \beta_{k} a_{k}
\end{align}
and
\begin{align}\label{eqn:14}
	\hat{a}_{k} \approx a_{k} + \beta_{k} a_{0}
	\,,
\end{align}
where the mixing angles $\beta_{k}$ are
\begin{align}\label{eqn:15}
	\beta_{k} = \frac{\overline{m}_{N}^{2} O_{N0} O_{Nk}}{m_{a_{k}}^{2}}\,.
\end{align}
The corresponding mass eigenvalues are
\begin{align}\label{eqn:16}
	m_{\hat{a}_{0}}^{2} \approx \overline{m}_{N}^{2} O_{N0}^{2} 
\end{align}
and
\begin{align}\label{eqn:17}
	m_{\hat{a}_{k}}^{2} \approx m_{a_{k}}^{2}\,.
\end{align}
To a good approximation $\hat{a}_{0} \approx a_{0}$ and $\hat{a}_{k} \approx a_{k}$.

Due to the shift symmetry Eq.~\eqref{eqn:2}, the $\hat{a}_{0}$ scalar can couple to other clockwork scalars only as a result of its couplings and mass term in Eq.~\eqref{eqn:8}, which break the shift symmetry. Therefore its quartic couplings to the other clockwork scalars will be highly suppressed, with a factor of $O_{N0}$ for each $\hat{a}_{0}$ scalar.

\subsection{Freeze-in decay processes} 
\begin{figure}[h]
\centering
\includegraphics[width=0.4\textwidth, angle=0]{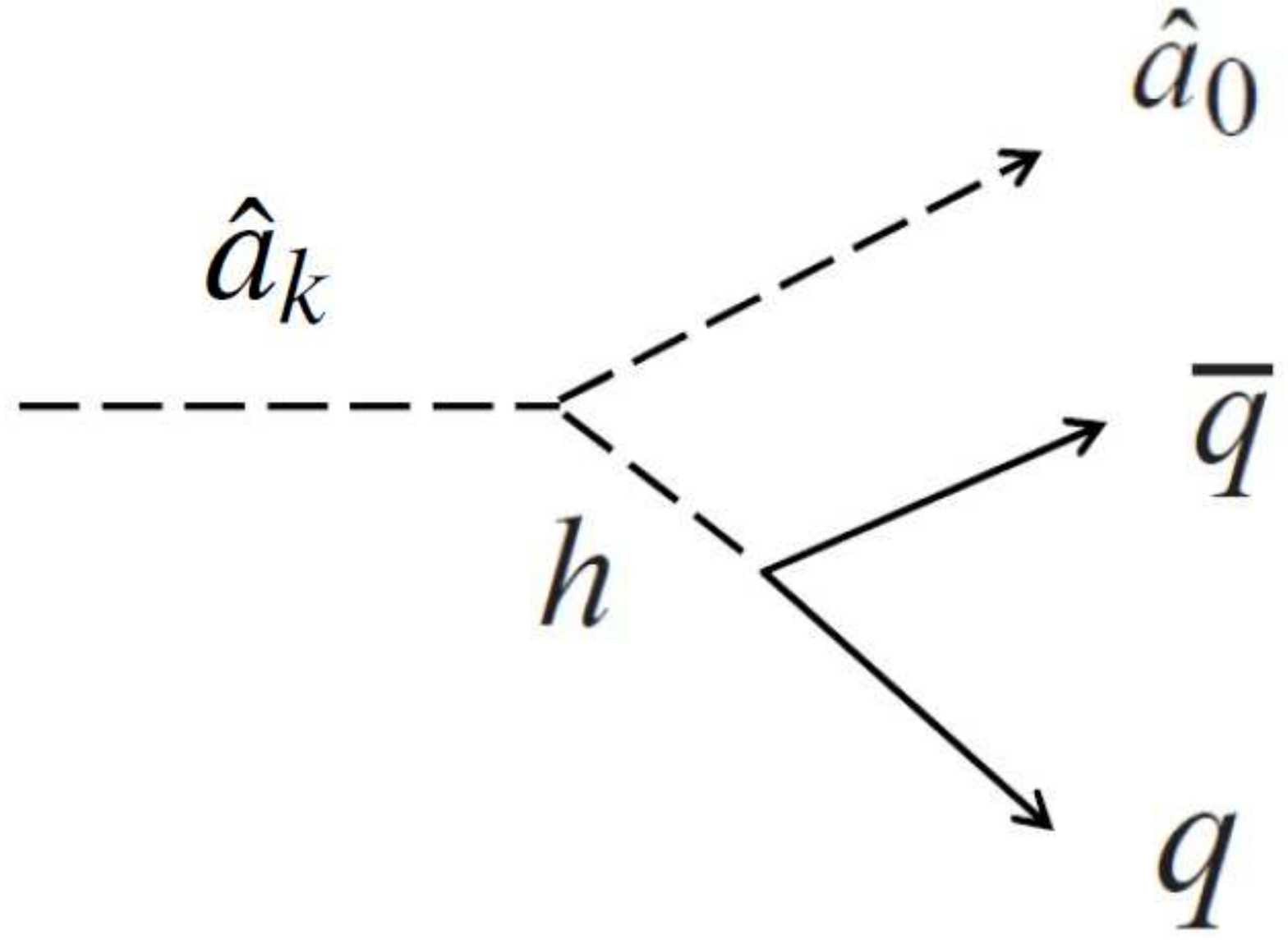} 
\caption{ Higgs exchange process for $\hat{a}_{k} \rightarrow \hat{a}_{0} + \overline{q}q$. }
\label{f1}
\end{figure}

There are two decay processes
\footnote{
There can also be scattering processes, via the quartic coupling of heavy clockwork scalars to $\hat{a}_{0}$, that could contribute to the freeze-in density of $\hat{a}_{0}$. As noted above, these couplings are suppressed by factors of $O_{N0}$ and are therefore no larger in magnitude than the coupling of $\hat{a}_{0}$ to the Higgs portal.  In general, for similar couplings, the freeze-in density will be dominated by decay processes over scattering processes \cite{jmfr,hall}.  However, as noted in \cite{goudelis}, since there are $N$ heavy clockwork scalars, the total contribution to the freeze-in density from scattering processes will be enhanced. In our analysis we are assuming that the quartic couplings $\lambda_{\pi_{j}}$ in Eq.~\eqref{eqn:1} and $\lambda_{\pi_{N}}$ in  Eq.~\eqref{eqn:8} are sufficiently small to counteract this enhancement of the quartic scattering processes,  such that the freeze-in density is dominated by decay processes. } which contribute to the freeze-in production of $\hat{a}_{0}$ dark matter: (i) $h \rightarrow \hat{a}_{k} + \hat{a}_{0}$ and (ii) $\hat{a}_{k} \rightarrow \hat{a}_{0} + \overline{f} f$, where $f = q\,,l$. The latter occurs via Higgs exchange, as shown in Fig.~\ref{f1} for the case $f=q$.
The corresponding decay rates are given by
\begin{align}\label{eqn:18}
	\Gamma_{h \rightarrow \hat{a}_{0} \hat{a}_{k}}^{TOT} \equiv \sum_{k} \Gamma_{h \rightarrow \hat{a}_{0} \hat{a}_{k}} = \frac{\lambda_{h \pi_{N}}^{2} v^2}{4 \pi m_{h}} O_{N0}^{2} \sum_{k}  O_{Nk}^{2}\,.
\end{align}
Since $O_{N0}^{2} + \sum_{k} O_{Nk}^{2} = 1$ and $O_{N0}^{2} \ll 1$, it follows that $\sum_{k} O_{Nk}^{2} \approx 1$. 
Therefore
\begin{align}\label{eqn:19}
	\Gamma_{h \rightarrow \hat{a}_{0} \hat{a}_{k}}^{TOT} = \frac{\lambda_{h \pi_{N}}^{2} v^2}{4 \pi m_{h}} O_{N0}^{2}\,.
\end{align}
For the second process we obtain for the 3-body decay process (for the case of final state quarks)
\begin{align}\label{eqn:20}
	\Gamma_{
	\hat{a}_{k} \rightarrow \hat{a}_{0}\overline{q}q}
	= \frac{3 \lambda_{h \pi_{N}}^{2} m_{q}^{2}
	m_{\hat{a}_{k}}^{3}O_{N0}^{2}O_{Nk}^{2}}{192 \pi^3 m_{h}^{4}} 
	\,,
\end{align}
where the factor of $3$ comes from the colour summation. Here we have assumed that $m_{\hat{a}_{k}} > 2m_{q}$.
We can estimate the total decay rate from summing over $k$ if we approximate the $\hat{a}_{k}$ as having a common mass,
\begin{align}\label{e21}
	\Gamma_{\hat{a}_{k} \rightarrow \hat{a}_{0}\overline{q}q
	}^{TOT} \equiv \sum_{k}  \Gamma_{
		\hat{a}_{k} \rightarrow \hat{a}_{0}\overline{q}q
	} = \frac{
		3 \lambda_{h \pi_{N}}^{2} O_{N0}^{2} m_{q}^{2}
		m_{\hat{a}_{k}}^{3} }{
		192 \pi^3 m_{h}^{4}}\,.
\end{align}
Comparing with $\Gamma_{h \rightarrow \hat{a}_{0} \hat{a}_{k}}^{TOT}$, since $m_{\hat{a}_{k}} < m_{h}$ and $m_{q} \ll m_{h}$ (the heaviest final state fermion being the b quark), we find that the $\hat{a}_{k}$ decay rate to $\hat{a}_{0}$ is negligible compared to the Higgs decay rate to $\hat{a}_{0}$. Therefore the freeze-in $\hat{a}_{0}$ dark matter density will be primarily due to Higgs decay.

\subsection{Freeze-in density} 
Freeze-in $\hat{a}_{0}$ dark matter is dominated by production of dark matter scalars via Higgs boson decay at $T \approx m_{h}$ \cite{jmfr,hall}. For $B_{1} \rightarrow B_{2}X$, where $X$ is the dark matter particle, the density from freeze-in is \cite{hall}
\begin{align}\label{eqn:22}
	\Omega_{X}h^{2} =
	\frac{ 1.1\times 10^{27} }{g_{*}^{3/2}}
	\frac{m_{X}\Gamma_{B_{1}}}{m_{B_{1}}^{2}}\,.
\end{align}
Thus from $h \rightarrow \hat{a}_{0} \hat{a}_{k}$, summed over all $k$, we obtain
\begin{align}\label{eqn:23}
	\Omega_{\hat{a}_{0}} h^2 = \frac{1.1 \times 10^{27}}{g_{*}^{3/2}} \frac{ \lambda_{h \pi_{N}}^{2} v^{2} O_{N0}^{2} m_{\hat{a}_{0}} }{4 \pi m_{h}^{3}}\,.
\end{align}
Using $m_{\hat{a}_{0}} = \overline{m}_{N} O_{N0}$, the dark matter density is therefore 
\begin{align}\label{eqn:24}
	\Omega_{\hat{a}_{0}} h^2 = \frac{1.1 \times 10^{27}}{g_{*}^{3/2}} \frac{ \lambda_{h \pi_{N}}^{2} v^{2} O_{N0}^{3} \overline{m}_{N} }{4 \pi m_{h}^{3}}\,.
\end{align}
Thus, to account for dark matter, we require that 
\begin{align}\label{eqn:25}
	O_{N0} = \left( g_{*}^{3/2} \frac{4 \pi m_{h}^{3} \Omega_{\hat{a}_{0}} h^2 }{ 1.1 \times 10^{27} \overline{m}_{N} \lambda_{h \pi_{N}}^{2} v^2 } \right)^{1/3} 
	=  1.2 \times 10^{-6}
	\left(\frac{1 \, {\rm GeV}}{\overline{m}_{N}} \right)^{1/3} \left( \frac{0.005}{\lambda_{h \pi_{N}}} \right)^{2/3}  \left(\frac{\Omega_{\hat{a}_{0}} h^{2}}{0.12} \right)^{1/3}
	\,,
\end{align}
where $g_{*} \approx 100$ for freeze-in Higgs decay at $T \approx m_{h}$.  
The corresponding $\hat{a}_{0}$ mass, from $m_{\hat{a}_{0}} = O_{N0} \overline{m}_{N} $, is  
\begin{align}\label{eqn:26}
	m_{\hat{a}_{0}} = 1.2 \, {\rm keV} \left(\frac{\overline{m}_{N}}{1 \, {\rm GeV}}\right)^{2/3} \left( \frac{0.005}{\lambda_{h \pi_{N}}} \right)^{2/3}
	\left(\frac{\Omega_{\hat{a}_{0}} h^{2}}{0.12} \right)^{1/3}
	\,.
\end{align}
In these we have normalized $\lambda_{h \pi_{N}}$ to a value which is consistent with the upper bound from the Higgs decay width, $\lambda_{h \pi_{N}} \lesssim 0.007$. Therefore $m_{\hat{a}_{0}} \sim 1-10 \, {\rm keV}$ is quite natural in this model.

From Eq.~\eqref{eqn:8}, the Higgs portal coupling responsible for freeze-in can be written as 
\begin{align}\label{eqn:27}
	\lambda_{1}\hat{a}_{0} \hat{a}_{k} |H|^{2}   \equiv 2 O_{N0} O_{Nk} \lambda_{h \pi_{N}}  \hat{a}_{0} \hat{a}_{k} |H|^{2}
	\,.
\end{align}
Thus from Eq.~\eqref{eqn:25} we find that
\begin{align}\label{eqn:28}
	\lambda_{1} = 1.2 \times 10^{-8} O_{Nk} \left(\frac{\overline{m}_{N}}{1 \, {\rm GeV}}\right)^{2/3} \left( \frac{0.005}{\lambda_{h \pi_{N}}} \right)^{5/3}
	\left(\frac{\Omega_{\hat{a}_{0}} h^{2}}{0.12} \right)^{1/3}
\end{align}

\begin{figure}
\centering
\includegraphics[width=0.5\textwidth, angle=0]{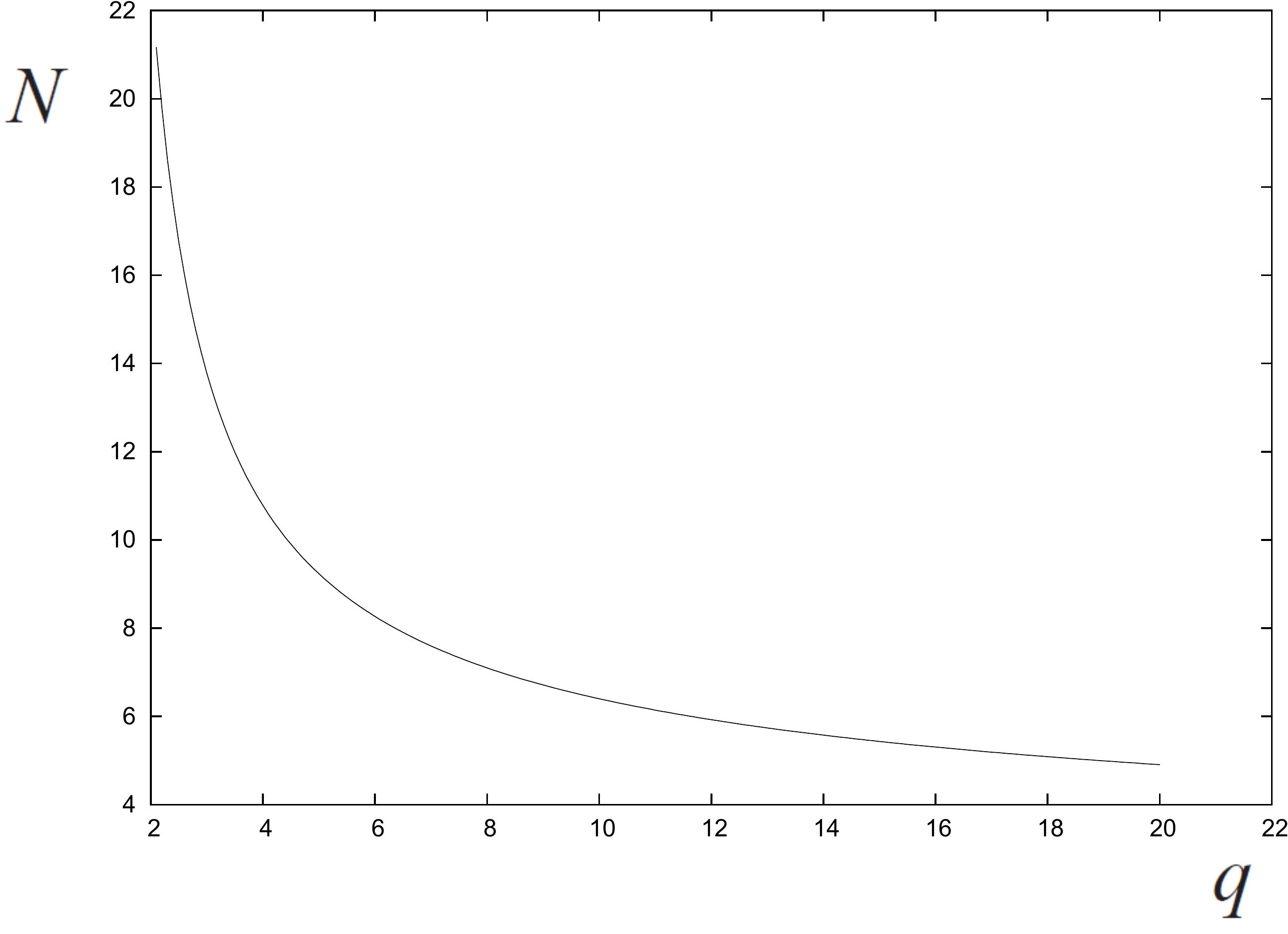} 
\caption{Values of $N$ versus $q$ necessary for freeze-in dark matter, for the case $\overline{m}_{N} = 10 \,{\rm GeV}$ and $\lambda_{h \pi_{N}} = 0.005$.}
\label{f2}
\end{figure}

The values of $q$ and $N$ which give the required value of $O_{N0}$ follow from Eq.~\eqref{eqn:4} and Eq.~\eqref{eqn:5}. 
Combining these with Eq.~\eqref{eqn:25} then gives
\begin{align}\label{eqn:29}
	N \ln q = \ln(\tilde{N}_{0}) - \ln(O_{N0}) = 
	13.6 + \ln(\tilde{N}_{0})  
	-\frac{1}{3} \ln \left(\frac{1 {\rm GeV}}{\overline{m}_{N}} \right)
	-\frac{2}{3} \ln \left(\frac{0.005}{\lambda_{h \pi_{N}}} \right)
	\,.
\end{align}
So $q$ and $N$ can be relatively small integers and still account for the necessary $O_{N0}$ to explain freeze-in dark matter via the Higgs portal. In Fig.~\ref{f2} we show that values of $N$ as a function of $q$ required to account for freeze-in dark matter for the case $\overline{m}_{N}=10\,{\rm GeV}$ and $\lambda_{h \pi_{N}} = 0.005$. For example, if $q = 4$ then $N \approx 10$ is required to account for freeze-in dark matter.

\section{Constraints on $\Lambda$ and $T_{R}$ from Freeze-in}
\label{sec:Constraints}
In order for $\hat{a}_{0}$ to play the role of dark matter produced by freeze-in, it is essential that its number density is much less than its thermal equilibrium value. The heavy clockwork scalars $\hat{a}_{k}$, $k = 1, ..., N$, do not have highly suppressed interactions with the Higgs boson and so will be in thermal equilibrium. The largest interaction of $\hat{a}_{0}$ with thermal particles will be via Eq.~\eqref{eqn:9}. This enables, for example, $\hat{a}_{0} + \hat{a}_{k} \leftrightarrow \hat{a}_{l} + \hat{a}_{m}$. We need to ensure that such processes do not result in a large density of $\hat{a}_{0}$ scalars.

As a representative example we will consider the non-renormalizable derivative interaction given by
\begin{align}\label{eqn:30}
	\frac{1}{\Lambda^4} \partial_{\mu} \pi_{1} \partial^{\mu} \pi_{1} \partial_{\nu} \pi_{1} \partial^{\nu} \pi_{1}
	\,.
\end{align}
This gives rise to the following interaction between $\hat{a}_{0}$ and $\hat{a}_{1}$ (using $a_{i} \approx \hat{a}_{i}$)
\begin{align}\label{eqn:31}
	\frac{O_{11}^{3} O_{10}}{\Lambda^4} \partial_{\mu} \hat{a}_{0} \partial^{\mu} \hat{a}_{1} \partial_{\nu} \hat{a}_{1} \partial^{\nu} \hat{a}_{1}\,.
\end{align}
Dimensionally, the scattering rate $\hat{a}_{0}\hat{a}_{1} \leftrightarrow \hat{a}_{1}\hat{a}_{1}$ is therefore 
\begin{align}\label{eqn:32}
	\Gamma \approx \frac{T^{9}}{\Lambda^{8}} \left(O_{11}^{3} O_{10}\right)^{2}
	\,.
\end{align}

To ensure that scattering does not produce a large density of $\hat{a}_{0}$ scalars, we will require that $\Gamma \lesssim H$ for all $T$. The largest possible value of $H$ is its value at the end of inflation, $H_{I}$. Assuming the inflaton decay products are instantly thermalized, the temperature during the inflaton matter-dominated era is $T \approx (H M_{Pl} T_{R}^{2})^{1/4}$. Therefore the largest temperature occurs at $H = H_{I}$. Since $\Gamma \propto T^9$ and $H \propto T^4$ during the inflaton matter-dominated era, if $\Gamma < H$ is satisfied at $H_{I}$ then it will be satisfied at all lower temperatures. Requiring that $\Gamma \lesssim H$ is satisfied at $H = H_{I}$ then gives a lower bound on $\Lambda$ 
\begin{align}\label{eqn:33}
	\Lambda \gtrsim (O_{11}^{3} O_{10})^{1/4} H_{I}^{5/32} M_{Pl}^{9/32} T_{R}^{9/16}\,.
\end{align}
 Therefore
\begin{align}\label{eqn:34}
	\Lambda \gtrsim  4.1 \times 10^{5} \,\, {\rm TeV} \times (O_{11}^{3} O_{10})^{1/4} \left( \frac{H_{I}}{10 ^{10}\, {\rm GeV}} \right)^{5/32} 
	\left( \frac{T_{R}}{1 \, {\rm TeV}} \right)^{9/16}
	\,.
\end{align}
Thus $\hat{a_{0}}$ production of heavy clockwork scalars by thermal scattering can be suppressed by assuming a sufficiently large value of $\Lambda$. In particular, Eq.~\eqref{eqn:34} is generally satisfied when $\Lambda \gtrsim M_{Pl}$, as suggested by naturalness of the SM.

\section{Collider phenomenology}
\label{sec:Collider}

In the case where at least some of the heavy clockwork scalars have mass less than $m_{h}/2$, it is possible for the Higgs boson to decay into pairs of heavy clockwork scalars. In this section we will consider the possible signatures of such a decay process.  

\subsection{Higgs decay width bound on the Higgs portal coupling}
We first consider the constraint on $\lambda_{h \pi_{N}}$ from the Higgs decay width. The decay rate for $h \rightarrow \hat{a}_{i} \hat{a}_{j}$ is
\begin{align}\label{eqn:35}
	\Gamma_{h \rightarrow \hat{a}_{i} \hat{a}_{j} } = \frac{\lambda_{h \pi_{N}}^{2} v^{2} O_{Ni}^{2} O_{Nj}^{2}} { 4 \pi m_{h}}
	\,.
\end{align}
Summing over $i$ and $j$ gives
\begin{align}\label{eqn:36}
	\Gamma_{h \rightarrow \hat{a}_{i} \hat{a}_{j} }^{TOT} = \sum_{i,j} \Gamma_{ h \rightarrow \hat{a}_{i} \hat{a}_{j} }  = \frac{\lambda_{h \pi_{N}}^{2} v^{2}}{4 \pi m_{h}}
	\,.
\end{align}
The 2-$\sigma$ upper bound on the branching ratio for Higgs decay to invisible states is \cite{atlas}
\begin{align}\label{eqn:37}
	BR_{inv} = \frac{\Gamma_{inv}}{\Gamma_{inv} + \Gamma_{SM}} < 0.28
	\,,
\end{align}
where $\Gamma_{SM} = 4.1 \, {\rm MeV}$ is the SM Higgs width.
Therefore 
\begin{align}\label{eqn:38}
	R_{inv} \equiv \frac{\Gamma_{inv}}{\Gamma_{SM}} < 0.39
	\,.
\end{align}
From Eq.~\eqref{eqn:36} and using $\Gamma_{inv} = \Gamma_{h \rightarrow \hat{a}_{i} \hat{a}_{j} }^{TOT}$, we then obtain a 2-$\sigma$ upper bound on $\lambda_{h\pi_{N}}$,
\begin{align}\label{eqn:39}
	\lambda_{h \pi_{N}}  <  \left(\frac{4 \pi m_{h}}{v^{2}} \right)^{1/2} \Gamma_{SM}^{1/2}
	\,R_{inv}^{1/2}
	\lesssim 0.007
	\,.
\end{align}

\subsection{An example of a sub-Higgs mass clockwork sector} 
\begin{figure}
\centering
\includegraphics[width=0.8\textwidth, angle=0]{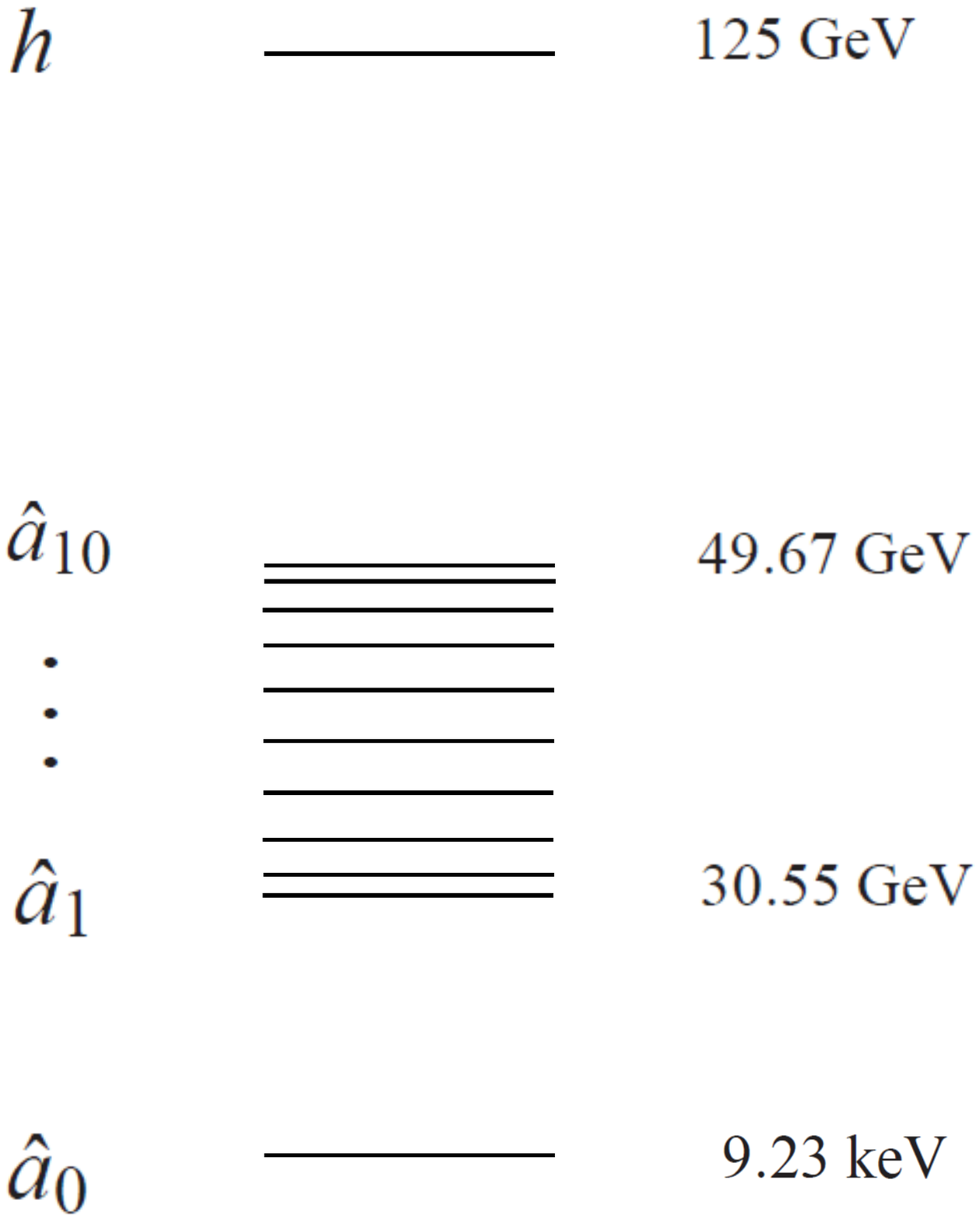}
\caption{Schematic illustration of the mass spectrum of the sub-Higgs mass scalar clockwork sector with $q = 4$, $N =10$ and $m=\overline{m}_{N} = 10$ GeV.}
\label{f3}
\end{figure}

To illustrate the class of model we are interested in, we will consider a clockwork sector with $N = 10$, $q = 4$ and $m = \overline{m}_{N} = 10 \, {\rm GeV}$. In order to account for the observed dark matter density, we then require (from Eq.~\eqref{eqn:29}) that $\lambda_{h \pi_{N}} = 0.0025$. With these inputs, we obtain the values of $\lambda_{k}$ and the corresponding heavy clockwork sector masses as given in Table 1 (using $m_{\hat{a}_{k}} \approx m_{a_{k}}$ for $k = 1, ... , N$). The mass of the $\hat{a}_{0}$ scalar is $m_{\hat{a}_{0}} = O_{N0} \overline{m}_{N} = 9.23$ keV (where $O_{N0} = 9.23 \times 10^{-7}$). The spectrum is illustrated schematically in Fig.~\ref{f3}.
\begin{table}[h]
\begin{center}
\begin{tabular}{|c|c|c|}
\hline $k$ & $\lambda_{k}$	& $m_{\hat{a}_{k}}$ (GeV) \\
\hline 1 &	$9.33$	&	$30.55$	 \\
\hline 2 &	$10.27$	&	$32.05$	 \\
\hline 3 &	$11.76$	&	$34.29$	 \\
\hline 4 &	$13.68$	&	$36.99$     \\
\hline 5 &	$15.86$	&	$39.82$	 \\
\hline 6 &	$18.14$	&	$42.60$	 \\
\hline 7 &	$20.32$	&	$45.08$	 \\
\hline 8 &	$22.24$	&	$47.16$	 \\
\hline 9 &	$23.73$	&	$48.71$	 \\
\hline 10 & $24.67$    &	$49.67$	 \\
\hline     
\end{tabular} 
\caption{ Masses of the heavy clockwork scalars for a model with $q = 4$, $N =10$ and $m=\overline{m}_{N}=10$ GeV. The corresponding mass for the dark matter scalar is $m_{\hat{a}_{0}} = 9.23$ keV. }
\end{center}
\end{table}

\subsection{Production of quark and lepton particle-antiparticle pairs plus missing energy as a signature of the clockwork sector} 
\begin{figure}
\centering
\includegraphics[width=0.7\textwidth, angle=0]{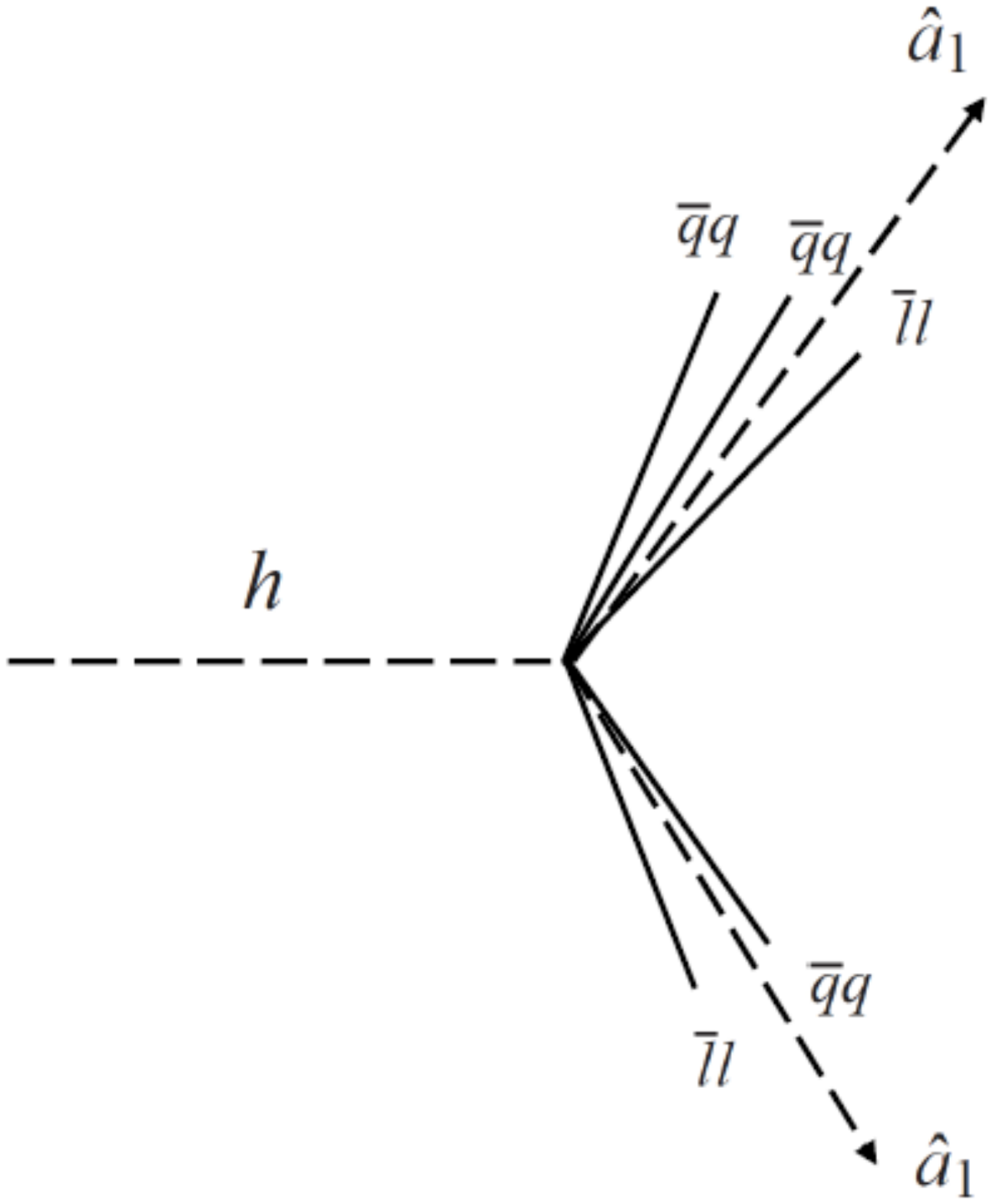}
\caption{An illustration of the decay of the Higgs boson into heavy clockwork scalars. The characteristic signature is prompt production of quark and lepton particle-antiparticle pairs plus missing energy in a pair of $\hat{a}_{1}$ scalars.}
\label{f4}
\end{figure}

Clockwork Higgs portal freeze-in models can have a distinctive Higgs decay phenomenology, which might become observable as the bound on the Higgs decay width improves. The Higgs can initially decay to clockwork scalars via $h \rightarrow \hat{a}_{i} \hat{a}_{j}$, with a branching ratio dependent upon the product $O_{Ni}O_{Nj}$.  
The $\hat{a}_{j}$ can then decay to the lightest scalar of the heavy clockwork scalar, $\hat{a}_{1}$, via a chain of intermediate Higgs exchange decay processes of the form  $\hat{a}_{j} \rightarrow \hat{a}_{k}\overline{f}f$, $\hat{a}_{k} \rightarrow \hat{a}_{l}\overline{f}f$, ..., $\hat{a}_{m} \rightarrow \hat{a}_{1}\overline{f}f$, where $f = q$ or $l$.  
This results in a characteristic Higgs decay signature, with a rapid production of different numbers of light quark and lepton particle-antiparticle pairs, depending upon the initial $\hat{a}_{j}$ pair and the subsequent decay chains, together with a long-lived $\hat{a}_{1}$ pair. This is illustrated in Fig.~\ref{f4}.

The $\hat{a}_{1}$ pairs produced from this process can only decay to $\hat{a}_{0} + \overline{q}q$ or $\overline{l}l$.
The decay rate for the process $\hat{a}_{1} \rightarrow \hat{a}_{0} \overline{f} f$, where $f = q$ or $l$, is
\begin{align}\label{eqn:40}
	\Gamma_{
		\hat{a}_{1} \rightarrow \hat{a}_{0}\overline{f}f
	} \approx \left(\frac{2 m_{\hat{a}_{1}}}{m_{h}}\right) \times  \frac{
		\lambda_{h \pi_{N}}^{2} O_{N0}^{2}O_{N1}^{2}m_{f}^{2}
		m_{\hat{a}_{1}}^{3}   }{
		192 \pi^3 m_{h}^{4}} \,,
\end{align}
with an additional factor of 3 for decay to quarks. In this we have included a time dilation factor $m_{\hat{a}_{1}}/E_{\hat{a}_{1}}$ with $E_{\hat{a}_{1}} \approx m_{h}/2$. Therefore the $\hat{a}_{1}$ lifetime is 
\begin{align}\label{eqn:41}
\tau_{\hat{a}_{1}} = 38.3 \, {\rm s} \times 
	\left( \frac{0.1}{O_{N1}}\right)^{2}
	\left( \frac{10^{-6}}{O_{N0}}\right)^{2}
	\left( \frac{0.005}{\lambda_{h \pi_{N}} }\right)^{2}
	\left( \frac{1 \, {\rm GeV}}{m_{f}}\right)^{2}
	\left( \frac{50 \, {\rm GeV}}{m_{\hat{a}_{1}}}\right)^{4}
	\,.
\end{align}
Therefore the $\hat{a}_{1}$ will escape the detector as missing energy.

Thus the signature of the production of clockwork sector scalars via Higgs decay will be prompt production of a distribution of quark and lepton particle-antiparticle pairs (with different numbers of quark and lepton pairs depending on the clockwork scalar decay chain to $\hat{a}_{1}$), accompanied by missing energy in the form of a pair of $\hat{a}_{1}$ scalars. The distribution of quark and lepton pairs will depend on the mixing angles $O_{Nj}$, which are determined by the clockwork parameters $q$ and $N$, and the mass of the clockwork sector scalars. Therefore it should be possible to determine whether such a decay process, if observed, is due to a light sector of clockwork scalars and to deduce the parameters $q$ and $N$ of the clockwork model.

\section{Nucleosynthesis and Warm Dark Matter Constraints}
\label{sec:NWDM}

A density of relic $\hat{a}_{1}$ scalars will be produced by thermal freeze-out. It is therefore important to determine the conditions under which the $\hat{a}_{1}$ scalars can decay rapidly enough to be consistent with primordial nucleosynthesis. The primary decay mode will be $\hat{a}_{1} \rightarrow \hat{a}_{0} + \overline{b}b$.  Primordial nucleosynthesis requires that particles decaying to $b$-quarks have a lifetime $\tau_{\hat{a}_{1}}  < 0.1$s \cite{kohri}. Since the decay of thermal relic $\hat{a}_{1}$ particles is non-relativistic, we can calculate the decay rate using Eq.~\eqref{eqn:40} but without the time-dilation factor and including the colour factor. The resulting lifetime is given by (using $m_{b} = 4.18$ GeV)
\begin{align}\label{eqn:41a}
	\tau_{\hat{a}_{1}} =  0.58 \, {\rm s} \times 
	\left( \frac{0.1}{O_{N1}}\right)^{2}
	\left( \frac{10^{-6}}{O_{N0}}\right)^{2}
	\left( \frac{0.005}{\lambda_{h \pi_{N}} }\right)^{2}
	\left( \frac{50 \, {\rm GeV}}{m_{\hat{a}_{1}}}\right)^{3}
	\,.
\end{align}
A more useful form for this expression is obtained by substituting $O_{N0}$ from Eq.~\eqref{eqn:25} and $O_{N1}$ from Eq.~\eqref{eqn:4}, and expressing $\overline{m}_{N}$ in terms of $m_{\hat{a}_{0}}$ via Eq.~\eqref{eqn:26}. This gives
\begin{align}\label{e41b}
	\tau_{\hat{a}_{1}} =  0.95 \, {\rm s} \times 
	\left( \frac{0.16}{O_{N1}}\right)^{2}
	\left( \frac{50 \, {\rm GeV}}{m_{\hat{a}_{1}}}	\right)^{3}
	\left( \frac{m_{\hat{a}_{0}}}{7 \, {\rm keV}}\right)
	\left(\frac{0.12}{\Omega_{\hat{a}_{0}} h^{2}} 	\right)
	\,,
\end{align}
where we have normalized $O_{N1}$ to its value for $q = 4$ and $N = 10$.  Thus for $q = 4$ and $N = 10$, the primordial nucleosynthesis constraint, $\tau_{\hat{a}_{1}} < 0.1$s, requires that $m_{\hat{a}_{0}} \lesssim 0.8$ keV  if $m_{\hat{a}_{1}} \lesssim 50 $ GeV, as expected when on-shell clockwork phenomenology via Higgs decay is possible.

The $\hat{a}_{1}$ scalars are mostly produced by the decay of Higgs bosons at $T \sim m_{h}$, therefore they will have approximately thermal energies when produced. Studies of Lyman-$\alpha$ constraints on conventional thermal warm dark matter (WDM), corresponding to dark matter fermions that decouple from equilibrium at a high temperature and are diluted to the observed dark matter density by a large entropy factor, estimate the lower bound on the mass of the WDM particle to be in the range 3-3.5 keV \cite{lyman, lyman2, lyman3}. Because the $\hat{a}_{1}$ scalars are produced at a lower temperature, with a correspondingly smaller entropy suppression, the temperature and energy of the resulting WDM will be higher, requiring a larger mass to satisfy the Lyman-$\alpha$ constraint. To estimate the lower bound in the case of clockwork scalar WDM produced via two-body Higgs boson decay, we will compare with the bound on freeze-in axino dark matter due to two-body decay of a heavy scalar \cite{axino}. In that case, the lower bound on the WDM particle mass corresponding to 3 keV for conventional WDM is found to be 7.3 keV. We will therefore consider a Lyman-$\alpha$ lower bound $m_{\hat{a}_{0}} > 7$ keV in the following.

Comparing with Eq.~\eqref{e41b} in the case $q = 4$ and $N = 10$, we find that it is not possible to satisfy the  Lyman-$\alpha$ lower bound on $m_{\hat{a}_{0}}$ and the nucleosynthesis constraint simultaneously when $m_{\hat{a}_{1}} \lesssim 50 $ GeV. This conclusion is generally true for all reasonably small values of $q$ and $N$. This excludes on-shell Higgs decay phenomenology in the case of a minimal SM sector, although off-shell Higgs decay to clockwork scalars would still be possible. The Lyman-$\alpha$ and nucleosynthesis constraints can still be satisfied with a sub-Higgs mass clockwork sector in the case where $m_{\hat{a}_{1}} > m_{h}/2$. For example, for $q = 4$ and $N = 10$, it is possible to have $m_{\hat{a}_{0}} > 7$ keV and $\tau_{\hat{a}_{1}} < 0.1$s if $m_{\hat{a}_{1}} > 106$ GeV. In this case the $\hat{a}_{0}$ dark matter could be warm enough to leave an observable effect on perturbation growth and Lyman-$\alpha$ observations.

Observable on-shell Higgs decays may be possible in simple extensions of the SM which have light singlet scalars. For example, in \cite{ram}, an extension is considered with an unstable light real singlet scalar. In this case, a portal-like coupling between $\pi_{N}^2$ and the light scalars would be expected, which could allow a much more rapid decay of $\hat{a}_{1}$ than the three-body decay to $b$-quarks in the SM case. This would allow the nucleosynthesis constraint to be satisfied with $m_{\hat{a}_{0}} > 7$ keV. In this case on-shell Higgs decay to clockwork scalars at colliders may be accompanied by  observable Higgs decay to light singlet scalars.

\section{Conclusions}
\label{sec:Conclusion} 

We have applied the clockwork mechanism to the freeze-in production of scalar dark matter via the Higgs portal, in the case where the clockwork scalars have masses less than the Higgs mass. This sub-Higgs mass case is particularly interesting from the point of view of testing the clockwork mechanism, as it could have a distinctive Higgs decay phenomenology which might be observable in collider experiments.

We have considered a minimal implementation of the clockwork mechanism, in which the clockwork sector has purely renormalizable interactions. In this case the cut-off scale of the effective renormalizable clockwork theory, $\Lambda$, can be arbitrarily large relative to the mass scale of the clockwork sector. This allows the dark matter scalars to evade thermalization due to possible non-renormalizable derivative interactions which are compatible with the shift symmetry of the clockwork sector.
A renormalizable clockwork sector with a mass scale of the order of the weak scale is also consistent with the idea that the electroweak scale is natural if there are no particles with masses between the weak scale and the Planck scale. In this case the clockwork sector and the Standard Model sector may be considered to have a common origin in a UV completion at a scale $\Lambda \gtrsim M_{Pl}$. 

In the case where at least some of the heavy clockwork scalars have mass less than $m_{h}/2$, the model has a characteristic on-shell Higgs decay phenomenology, with the Higgs first decaying to a pair of heavy clockwork scalars which subsequently decay, via a chain of decays to lighter clockwork scalars, to multiple particle-antiparticle pairs of quarks and leptons plus missing energy in the form of a pair of very long-lived next-to-lightest clockwork scalars. Observation of Higgs decays to a distribution of quark and lepton particle-antiparticle pairs plus missing energy could therefore allow the clockwork origin of the decays to be confirmed and the parameters of the clockwork sector, $q$ and $N$, to be determined. 

Primordial nucleosynthesis constraints on the lifetime of the $\hat{a}_{1}$ scalar and Lyman-$\alpha$ constraints on the mass of the $\hat{a}_{0}$ dark matter scalar impose strong constraints on the model, which exclude the case of a minimal Standard Model observable sector if the heavy clockwork scalars are light enough to be produced via on-shell Higgs decay. 
The cosmological constraints can still be satisfied with a heavier (but still sub-Higgs mass) clockwork sector. In this case the dark matter particles may be warm enough to have a significant effect on perturbation growth and Lyman-$\alpha$ observations. Production of clockwork scalars via off-shell Higgs decays would also be possible in this case. This would offer the best prospect for testing the model at the LHC when the clockwork sector is coupled to the minimal SM.

Observable on-shell Higgs decay to clockwork scalars may be possible in simple extensions of the Standard Model with light singlet scalars, which allow the $\hat{a}_{1}$ scalars to decay quickly enough to the light scalars to satisfy the nucleosynthesis constraint. In this case it may be possible to produce both clockwork scalars and light singlet scalars via on-shell Higgs decay at colliders.

\section*{Note Added}  Recently, a new paper on clockwork freeze-in models appeared on arXiv \cite{goudelis}, which presents a more general analysis of clockwork freeze-in models for the case of a 1-100 TeV clockwork sector.

\section*{Acknowledgements} The work of JM was partially supported by STFC via the Lancaster-Manchester-Sheffield Consortium for Fundamental Physics.



\begin{thebibliography}{99}





\bibitem{choi}
K.~Choi and S.~H.~Im,
JHEP {\bf 1601}, 149 (2016)
doi:10.1007/JHEP01(2016)149
[arXiv:1511.00132 [hep-ph]].



\bibitem{cl1}
D.~E.~Kaplan and R.~Rattazzi,
Phys.\ Rev.\ D {\bf 93}, no. 8, 085007 (2016)
doi:10.1103/PhysRevD.93.085007
[arXiv:1511.01827 [hep-ph]].
  
\bibitem{giudice}
G.~F.~Giudice and M.~McCullough,
JHEP {\bf 1702}, 036 (2017)
doi:10.1007/JHEP02(2017)036
[arXiv:1610.07962 [hep-ph]].


\bibitem{CWgeneral}
I.~Ben-Dayan,
arXiv:1706.05308 [hep-ph].

\bibitem{CWgauged}
H.~M.~Lee,
Phys.\ Lett.\ B {\bf 778}, 79 (2018)
doi:10.1016/j.physletb.2018.01.010
[arXiv:1708.03564 [hep-ph]].
  
\bibitem{CWDM}
T.~Hambye, D.~Teresi and M.~H.~G.~Tytgat,
JHEP {\bf 1707}, 047 (2017)
doi:10.1007/JHEP07(2017)047
[arXiv:1612.06411 [hep-ph]].

\bibitem{CWneutrino}
S.~C.~Park and C.~S.~Shin,
Phys.\ Lett.\ B {\bf 776}, 222 (2018)
doi:10.1016/j.physletb.2017.11.057
[arXiv:1707.07364 [hep-ph]].

\bibitem{CWmuong2}
D.~K.~Hong, D.~H.~Kim and C.~S.~Shin,
Phys.\ Rev.\ D {\bf 97}, no. 3, 035014 (2018)
doi:10.1103/PhysRevD.97.035014
[arXiv:1706.09376 [hep-ph]].

\bibitem{CWaxions}
R.~Coy, M.~Frigerio and M.~Ibe,
JHEP {\bf 1710}, 002 (2017)
doi:10.1007/JHEP10(2017)002
[arXiv:1706.04529 [hep-ph]];
M.~Farina, D.~Pappadopulo, F.~Rompineve and A.~Tesi,
JHEP {\bf 1701}, 095 (2017)
doi:10.1007/JHEP01(2017)095
[arXiv:1611.09855 [hep-ph]];
A.~J.~Long,
arXiv:1803.07086 [hep-ph].
  
\bibitem{CWComposite}
A.~Ahmed and B.~M.~Dillon,
Phys.\ Rev.\ D {\bf 96}, no. 11, 115031 (2017)
doi:10.1103/PhysRevD.96.115031
[arXiv:1612.04011 [hep-ph]].

\bibitem{wgc}
P.~Saraswat,
Phys.\ Rev.\ D {\bf 95}, no. 2, 025013 (2017)
doi:10.1103/PhysRevD.95.025013
[arXiv:1608.06951 [hep-th]].


\bibitem{CWinflation} 
A.~Kehagias and A.~Riotto,
Phys.\ Lett.\ B {\bf 767}, 73 (2017)
doi:10.1016/j.physletb.2017.01.042
[arXiv:1611.03316 [hep-ph]];
S.~H.~Im, H.~P.~Nilles and A.~Trautner,
JHEP {\bf 1803}, 004 (2018)
doi:10.1007/JHEP03(2018)004
[arXiv:1707.03830 [hep-ph]].


\bibitem{craig}
N.~Craig, I.~Garcia Garcia and D.~Sutherland,
JHEP {\bf 1710}, 018 (2017)
doi:10.1007/JHEP10(2017)018
[arXiv:1704.07831 [hep-ph]].

\bibitem{reply}
G.~F.~Giudice and M.~McCullough,
arXiv:1705.10162 [hep-ph].

\bibitem{jmfr}
J.~McDonald,
Phys.\ Rev.\ Lett.\  {\bf 88}, 091304 (2002)
doi:10.1103/PhysRevLett.88.091304
[hep-ph/0106249].

\bibitem{hall}
L.~J.~Hall, K.~Jedamzik, J.~March-Russell and S.~M.~West,
JHEP {\bf 1003}, 080 (2010)
doi:10.1007/JHEP03(2010)080
[arXiv:0911.1120 [hep-ph]].

\bibitem{tenkanen} 
N.~Bernal, M.~Heikinheimo, T.~Tenkanen, K.~Tuominen and V.~Vaskonen,
Int.\ J.\ Mod.\ Phys.\ A {\bf 32}, no. 27, 1730023 (2017)
doi:10.1142/S0217751X1730023X
[arXiv:1706.07442 [hep-ph]].

\bibitem{Heikinheimo:2016yds} 
  M.~Heikinheimo, T.~Tenkanen, K.~Tuominen and V.~Vaskonen,
  Phys.\ Rev.\ D {\bf 94}, no. 6, 063506 (2016)
  Erratum: [Phys.\ Rev.\ D {\bf 96}, no. 10, 109902 (2017)]
  doi:10.1103/PhysRevD.96.109902, 10.1103/PhysRevD.94.063506
  [arXiv:1604.02401 [astro-ph.CO]].
  
\bibitem{Heikinheimo:2017ofk} 
  M.~Heikinheimo, T.~Tenkanen and K.~Tuominen,
  Phys.\ Rev.\ D {\bf 96}, no. 2, 023001 (2017)
  doi:10.1103/PhysRevD.96.023001
  [arXiv:1704.05359 [hep-ph]].
  
\bibitem{Enqvist:2017kzh} 
  K.~Enqvist, R.~J.~Hardwick, T.~Tenkanen, V.~Vennin and D.~Wands,
  JCAP {\bf 1802}, no. 02, 006 (2018)
  doi:10.1088/1475-7516/2018/02/006
  [arXiv:1711.07344 [astro-ph.CO]].

\bibitem{noscale}
M.~Shaposhnikov,
arXiv:0708.3550 [hep-th].


\bibitem{agrav}
A.~Salvio and A.~Strumia,
Eur.\ Phys.\ J.\ C {\bf 78}, no. 2, 124 (2018)
doi:10.1140/epjc/s10052-018-5588-4
[arXiv:1705.03896 [hep-th]].


\bibitem{chp1}
  J.~Kim and J.~McDonald,
  Phys.\ Rev.\ D {\bf 98}, 023533 (2018)
  doi:10.1103/PhysRevD.98.023533
  [arXiv:1709.04105 [hep-ph]].



\bibitem{ram}  V.~Barger, P.~Langacker, M.~McCaskey, M.~J.~Ramsey-Musolf and G.~Shaughnessy, 
Phys.\ Rev.\ D {\bf 77}, 035005 (2008)
doi:10.1103/PhysRevD.77.035005
[arXiv:0706.4311 [hep-ph]].

\bibitem{goudelis} A.~Goudelis, K.~A.~Mohan and D.~Sengupta,
  JHEP {\bf 1810}, 014 (2018)
  doi:10.1007/JHEP10(2018)014
  [arXiv:1807.06642 [hep-ph]].





\bibitem{atlas}
G.~Aad {\it et al.} [ATLAS Collaboration],
JHEP {\bf 1601}, 172 (2016)
doi:10.1007/JHEP01(2016)172
[arXiv:1508.07869 [hep-ex]].


\bibitem{kohri} M.~Kawasaki, K.~Kohri, T.~Moroi and Y.~Takaesu, 
Phys.\ Rev.\ D {\bf 97}, no. 2, 023502 (2018)
doi:10.1103/PhysRevD.97.023502
[arXiv:1709.01211 [hep-ph]].



\bibitem{lyman} M.~Viel, G.~D.~Becker, J.~S.~Bolton and M.~G.~Haehnelt, 
Phys.\ Rev.\ D {\bf 88}, 043502 (2013)
doi:10.1103/PhysRevD.88.043502
[arXiv:1306.2314 [astro-ph.CO]].
\bibitem{lyman2} J.~Baur, N.~Palanque-Delabrouille, C.~Yèche, C.~Magneville and M.~Viel, 
JCAP {\bf 1608}, no. 08, 012 (2016)
doi:10.1088/1475-7516/2016/08/012
[arXiv:1512.01981 [astro-ph.CO]].

\bibitem{lyman3}V.~Irsic {\it et al.}, 
Phys.\ Rev.\ D {\bf 96}, no. 2, 023522 (2017)
doi:10.1103/PhysRevD.96.023522
[arXiv:1702.01764 [astro-ph.CO]].

\bibitem{axino} K.~J.~Bae, A.~Kamada, S.~P.~Liew and K.~Yanagi, 
JCAP {\bf 1801}, no. 01, 054 (2018)
doi:10.1088/1475-7516/2018/01/054
[arXiv:1707.06418 [hep-ph]].




\end{thebibliography}
\end{document}